\documentclass[aps,prb,twocolumn,groupedaddress,showpacs]{revtex4}
\usepackage{graphicx}
\usepackage[colorlinks,citecolor=blue,urlcolor=blue, linkcolor=blue]{hyperref}
\usepackage{epstopdf}   

\graphicspath{{figures//}}

\begin{document}

\title{The nature of high-energy radiation damage in iron: Modeling results}
\author{E. Zarkadoula, M. T. Dove and K. Trachenko}
\address{SEPnet and School of Physics and Astronomy, Queen Mary University of London, Mile End Road, London, E1 4NS, UK}
\author{S. L. Daraszewicz and D. M. Duffy}
\address{London Centre for Nanotechnology, Department of Physics and Astronomy, University College London, Gower Street, London, WC1E 6BT, UK}
\author{M. Seaton and I. T. Todorov}
\address{Computational Science and Engineering Department, CCLRC Daresbury Laboratory, Keckwick Lane, Daresbury,
Warrington WA44AD, Cheshire, UK}
\author{K. Nordlund}
\address{University of Helsinki, P.O. Box 43, FIN-00014 Helsinki, Finland}

\begin{abstract}
Understanding and predicting a material's performance in response to high-energy radiation damage, as well as designing future materials to be used in intense radiation environments, requires the knowledge of the structure, morphology and amount of radiation-induced structural changes \cite{dud1,stoneham,dud2,weber,right}. We report the results of molecular dynamics simulations of high-energy radiation damage in iron in the range 0.2--0.5 MeV. We analyze and quantify the nature of collision cascades both at the global and local scale. We find that the structure of high-energy collision cascades becomes increasingly continuous as opposed to showing sub-cascade branching reported previously. At the local length scale, we find large defect clusters and novel small vacancy and interstitial clusters. These features form the basis for physical models aimed at understanding the effects of high energy radiation damage in structural materials.
\end{abstract}


\maketitle

\section{Introduction}

Radiation effects are common in nature. Their sources vary from cosmic radiation to decay of isotopes in terrestrial rocks. A large variety of radiation sources are also created and used in science and technology. This includes energy generation in existing nuclear power stations, where kinetic energy of fission products is converted into heat and electricity. When feasible, future fusion reactors will harvest the energy from thermonuclear reactions. In these applications, the energy of emitted particles has a two-fold effect: on one hand, this energy is converted into useful energy, by heating the material; on the other hand, this energy damages the material and degrades the properties important for the operation, including mechanical, thermal, transport and other properties. This is currently a central issue for fusion reactors, where the ability of metal structural components to withstand very high neutron fluxes is intensely discussed \cite{dud1,stoneham,dud2}. Another example is the damage to nuclear reactor materials coming from fission products. In addition, the nuclear industry faces yet another problem, that of radiation damage to materials to be used to encapsulate long-lived radioactive waste \cite{weber,synroc2,gei1}.

A heavy energetic particle displaces atoms on its path which, in turn, displace other atoms in the system. A collection of these atoms is often referred to as a ``collision cascade'' \cite{Ave98,ishino,k6,k1,k2,k3,k4}. A typical collision cascade created by a heavy 100 keV particle propagates and relaxes on the order of picoseconds and spans nanometers. The resulting structural damage, in the form of amorphous pockets or point defects and their clusters, ultimately defines to what extent materials' mechanical, thermal and other properties are altered. For example, radiation-induced defects can reduce materials' thermal conductivity and therefore result in inefficient energy transfer in both fusion and nuclear reactors, heat localization and other unwanted effects.

Understanding and predicting these and other effects, as well as designing future materials to be used in intense radiation environments, requires developing physical models. These, in turn, are based on the knowledge of what high-energy radiation damage is in terms of structure and morphology.

Molecular dynamics (MD) simulations have been an important method for studying radiation damage in materials because they give access to the small time and length scales of the collision cascades, and give a detailed picture of the damage at the atomistic scale. Previous MD simulations have provided important insights into the radiation damage process \cite{Ave98,ishino,k6,k1,k2,k3,k4,stoller2}. However, due to system size limitations in MD simulations, the reported results were limited to energies of about 100 keV. Recently, the number of displaced atoms was reported for 200 keV cascade \cite{st-rev}.

On the other hand, knock-on energies are larger in several important applications. When impacted on 14 MeV neutrons, iron knock-on atoms in fusion reactors reach the energy of up to 1 MeV \cite{stoneham,ishino,st-rev} with an average energy of about 0.5 MeV \cite{stoller2}. In fission nuclear reactions, the fission product energies are on the order of 50 to 100 MeV, transferring high energy to the surrounding material. The need to simulate realistic energy cascades has been particularly emphasized, with a view that extrapolation of low-energy results may not account for some important features of higher-energy radiation process which can contain novel qualitative features. More generally, the need to simulate length and energy scales that are relevant and appropriate to a particular physical process has been recognized and reiterated \cite{right}.

In this paper, we study the radiation damage process due to high-energy Fe knock-on atoms of 0.2--0.5 MeV energy. We focus on high-energy radiation damage in $\alpha$-iron, the main structural material in fusion and future fission reactors. We analyze and quantify the nature of collision cascades both at the global and local scale. We find that high-energy collision cascades may propagate and relax as increasingly continuous damage structures as opposed to showing sub-cascade branching as assumed previously. At the local length scale, we find large clusters and new defect structures.

\section{Methods}

We have used the DL$\_$POLY program, a general-purpose package designed for large-scale simulations \cite{dl1,dl2}. We have simulated systems with 100-500 million of atoms, and run MD simulations on 20000--60000 parallel processors of the HECToR National Supercomputing Service \cite{hector}. MD simulations were performed at a constant energy and volume ensemble (except for the boundary layer atoms, see below) with an initial temperature set to 300 K, which was preceded by equilibration runs in a constant pressure ensemble at 300 K. Periodic boundary conditions were imposed in all directions.

We have implemented several features to handle radiation damage simulations. First, we used a variable time step to account for faster atomic motion at the beginning of the cascade development and its gradual slowing down at later stages. Second, the MD box boundary layer of thickness of about 10 \AA\ was connected to a constant temperature thermostat at 300 K to emulate the effect of energy dissipation into the sample. Third, we have accounted for the electronic energy losses (EEL), particularly important at high energies. EEL is a complicated process that involves a wide range of effects affecting damage production and annealing \cite{stoneham,race,duffy1,caro}. Taking EEL into account in MD simulations involves models that include slowing-down of atoms due to energy transfer to electron excitation processes as well as feeding this energy back to the system as the phonon energy. The implementation of this dual energy exchange mechanism in MD simulations, based on the two-temperature approach \cite{ttm,duffy1,duffy2}, is in progress. In this work, we model electronic energy loss as a friction term added to the equations of motion. The characteristic energy loss relaxation time (taken here as $\tau_{es} = 1.0$ ps), is obtained by relating the stopping strength \mbox{($\lambda = 0.1093$ eV $^{1/2}$\AA$^{-1}$)\cite{zbl}} in the low-velocity regime via Lindhard's model to the rate of energy loss for a single atom \cite{lindhard,duffy2}. Such electronic stopping would only be effective above a certain threshold, where the atoms would have sufficient energy to scatter inelastically. We use a cut-off kinetic energy value (8.6 eV) corresponding to twice of the cohesive energy \cite{coh2}, however a number of other threshold values have been proposed \cite{coh1,coh3,coh4,coh5}.

For \mbox{$\alpha$-Fe}, we have used a many-body embedded-atom potential \cite{mendel}, optimized for better reproduction of several important properties of $\alpha$-Fe including the energetics of point defects and their clusters (``M07'' from Ref. \cite{malerba}). At distances shorter than 1 \AA, interatomic potentials were joined to short-range repulsive ZBL potentials \cite{zbl}.

To analyze the collision cascade, we have employed two methods. First, an atom is identified as ``displaced'' if it moves more than distance $d$ from its initial position. The number of displaced atoms, $N_{\mathrm{disp}}$, quantifies the overall amount of introduced damage. Some of this damage recovers back to crystal. To account for this effect, we employed the second method in which an atom is identified as a ``defect'' if it is further than distance $d$ from any of the crystalline position of the original lattice. Then, the number of defects $N_{\mathrm{def}}$ quantifies both damage production and its recovery. We note that $N_{\mathrm{disp}}$ and $N_{\mathrm{def}}$ depend on $d$ (we use $d=0.75$ \AA\ as a convenient measure). However, the trends discussed in Sec. \ref{results}, including the two regimes of cascade relaxation as well as dynamics of defects recovery are not sensitive to $d$ provided it is in the sensible range of distances (e.g., too small $d\lesssim 0.1-0.2$ \AA\ will be affected by usual thermal fluctuations whereas \mbox{$d\gtrsim 1-1.5$ \AA} may not identify defect atoms).

Vacancies or self-interstitial atoms (SIA) are defined to belong to the same defect group (cluster) if within 2 nearest-neighbour distance (plus a 20\% perturbation). Second nearest-neighbour (nn) is a common clustering criterion for SIAs \cite{st-rev, bjork}, however the criteria for vacancy clusters vary  significantly (from 1 to 4nn) across the literature \cite{malerba2}. When identifying cluster size in the \mbox{Sec. \ref{results}} the net defect number (the difference between the number of SIAs and the number of vacancies) is reported.

\section{Results and discussion} \label{results}

We discuss the main features of high-energy collision cascades. To account for potentially different collision cascades due to different knock-on directions, we have simulated 4 different directions for each energy, avoiding low-density directions and associated channeling.

\begin{figure}[thb]
\begin{flushleft}
\includegraphics[width=\columnwidth]{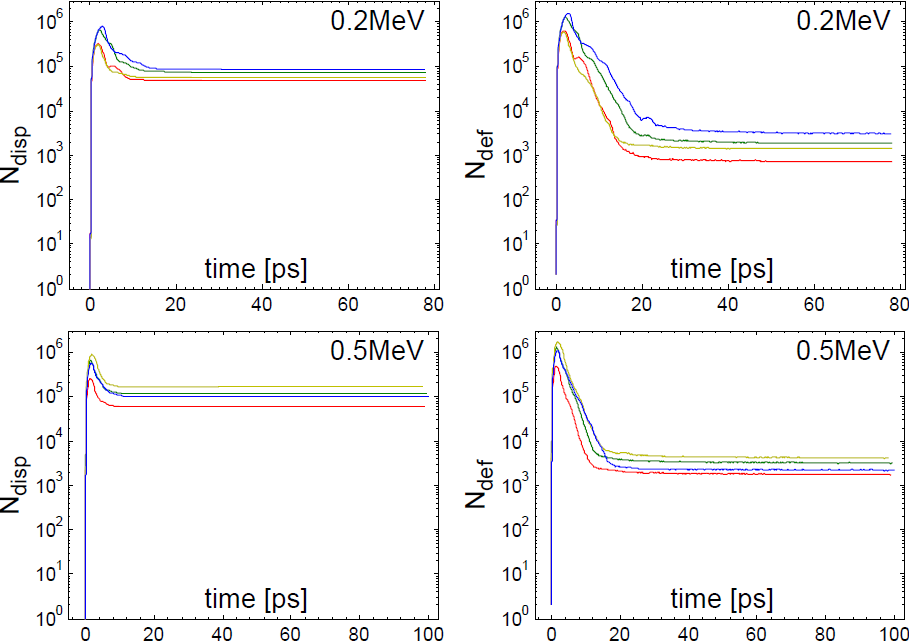}
\end{flushleft}
\caption{$N_{\mathrm{disp}}$ and $N_{\mathrm{def}}$ from 0.2 MeV (top) and 0.5 MeV (bottom) knock-on atoms for four cascades.}
\label{fig1}
\end{figure}

$N_{\mathrm{disp}}$ and $N_\mathrm{def}$ are shown in Fig. \ref{fig1} for 0.2 and \mbox{0.5 MeV} cascades simulated in different knock-on directions. We observe two types of cascade relaxation in Fig. \ref{fig1}. The first type is related to the large peak of $N_{\mathrm{disp}}\approx 10^6$ at short times of about 1--2 ps. This peak relaxes during about 10 ps. This peak is often discussed as the ``thermal spike'' \cite{spike,Sam07d} related to melting inside the collision cascade and associated swelling that causes the temporary increase of $N_{\mathrm{disp}}$. At the atomistic level, the increase of $N_{\mathrm{disp}}$ can be understood on the basis of anharmonicity of interatomic interactions: large-scale atomic motion inside the cascade causes the increase of interatomic separations due to anharmonicity. This results in the outward pressure from the cascade on the surrounding lattice and lattice expansion. This elastic deformation lasts several ps, equal to several periods of atomic vibrations during which the energy is dissipated to the lattice, and gives rise to the peak of $N_{\mathrm{disp}}$. Notably, the elastic deformation is reversible irrespective of whether it is followed by the recovery of the structural damage discussed below. In Fig. \ref{fig1}, $N_{\mathrm{disp}}$, averaged over all knock-on directions at the end of simulation (corresponding to the flat lines in Fig. \ref{fig1}) is about 67,000 and 111,000 atoms for 0.2 MeV and 0.5 MeV cascades, respectively.

\begin{figure}
\begin{center}
\includegraphics[width=\columnwidth]{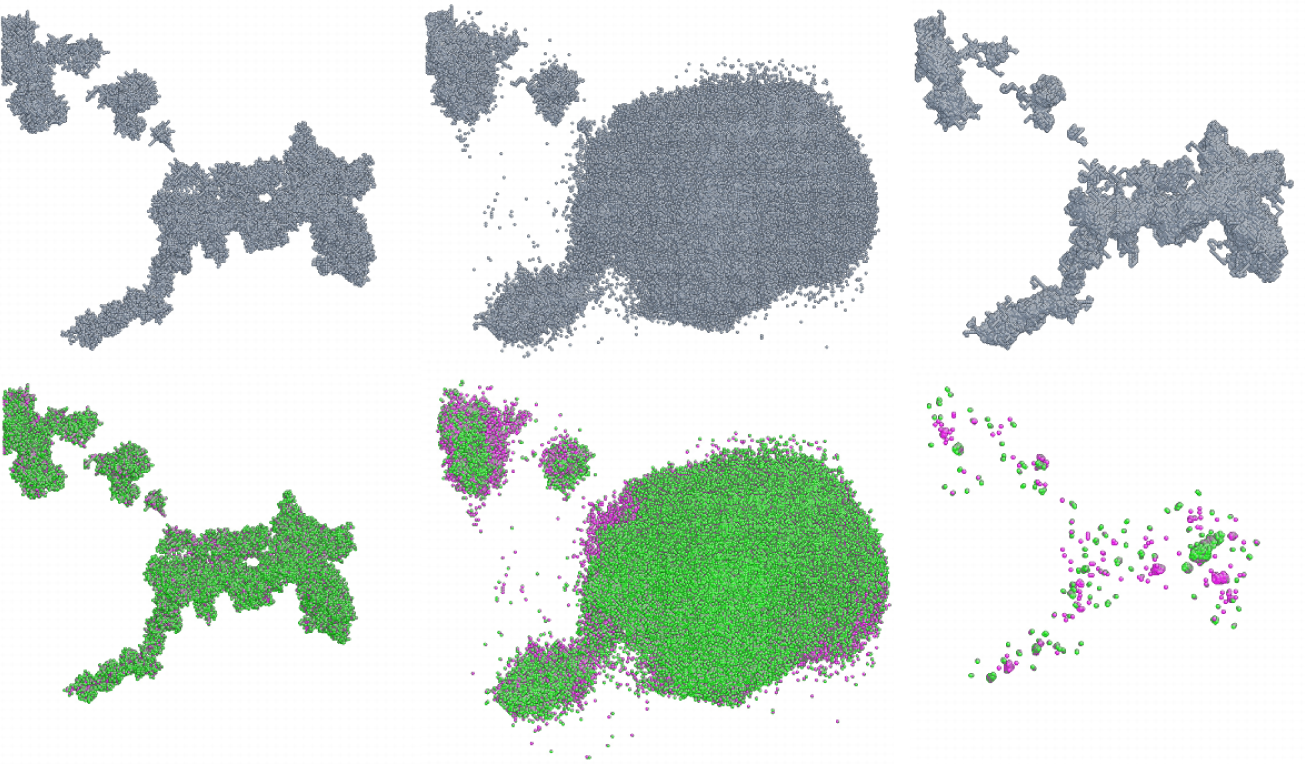}
\end{center}
\caption{Displaced (top) and defect (bottom) atoms in a representative 0.2 MeV collision cascade. The knock-on atom moves from the top left to the bottom right corner. The three frames for each type of atoms correspond to 0.3 ps, 3 ps and 80 ps, respectively. Cascade size (maximal separation between any two atoms in the cascade) is 560 \AA. Vacancies (interstitials) are represented in purple (green); we used Atomeye software \cite{atomeye} to visualize cascade evolution.
}
\label{fig2}
\end{figure}

\begin{figure}
\begin{center}
\includegraphics[width=\columnwidth]{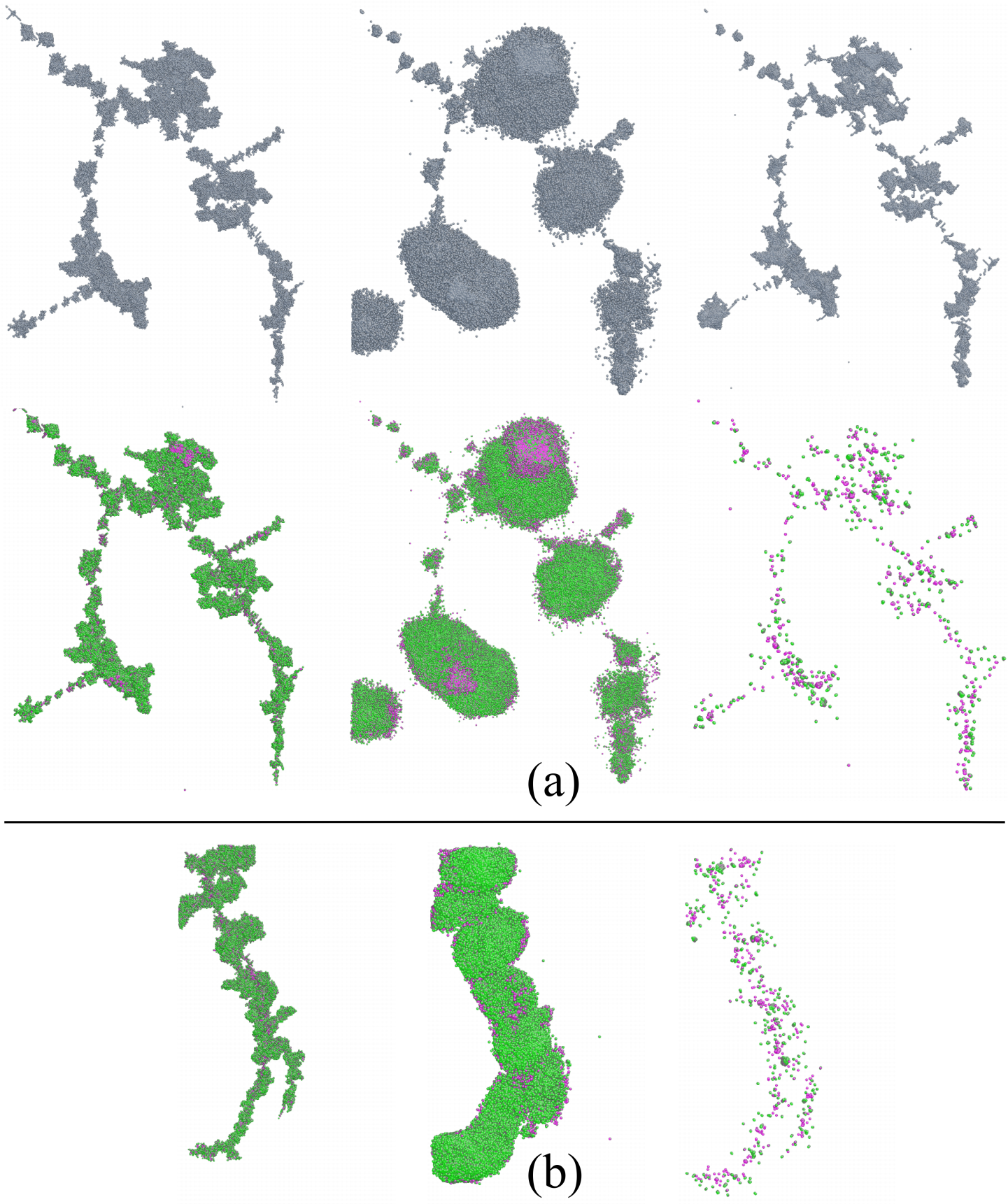}
\end{center}
\caption{(a) and (b) show two representative 0.5 MeV cascades. The knock-on atom moves from the top left to the bottom right corner. In (a) we show both displaced (top) and defect (bottom) atoms at 0.2, 1.5 ps and 100 ps. In (b) we show the defect atoms only at 0.3 ps, 2 ps and 100 ps. Cascade size in (a) and (b) is 950 \AA\ and 1300 \AA, respectively.
}
\label{fig3}
\end{figure}

Figures \ref{fig2}--\ref{fig3} show the snapshots of the collisions cascades at three different stages of cascade development that include the intermediate stage corresponding to the large peak of $N_{\mathrm{disp}}$ in Fig. \ref{fig1}. Consistent with \mbox{Fig. \ref{fig1}}, we observe a significantly larger number of atoms involved in large displacements at intermediate times as compared to the final relaxed state.

The second type of cascade relaxation is related to the dynamics of $N_{\mathrm{def}}$. At short ps times, the large peak of $N_{\mathrm{def}}$ is of the same origin as that seen for $N_{\mathrm{disp}}$. However, dynamics of $N_{\mathrm{def}}$ also reflects the recovery of structural damage. This recovery proceeds by the diffusion and recombination processes during which atoms settle at the newly found crystalline positions. This process lasts up to 20 ps, significantly longer than relaxation time of the first elastic relaxation process (see Fig. \ref{fig1}). As a result of this relaxation, $N_{\mathrm{def}}$, averaged over all simulated directions at the end of simulation (corresponding to the flat lines in Fig. \ref{fig1}), is about 1,800 and 2,800 atoms, respectively, corresponding to approximately 97\% recovery rate as compared to $N_{\mathrm{disp}}$. Such a high recovery rate is in interesting resemblance to some of the resistant ceramic materials, but in contrast to others \cite{prb,prb1}.

Our simulations provide an important insight about the structure and morphology of high-energy cascades. The existing view of the high-energy cascade is that it branches out to smaller sub-cascades and ``pockets'' of damage that are well separated from each other. This takes places over a certain energy threshold, even though this threshold was not firmly established \cite{Nor98,Hei94,st-rev}. This picture originated as a result of using binary-collision simulations in combination with MD simulations of low-energy events. Although involving approximations inherent in binary-collision simulations and extrapolations of low-energy MD simulations, this picture would offer a great degree of reduction and simplification: in analyzing the results and consequences of high-energy damage, only small sub-cascades need to be considered.

In Figures \ref{fig2}--\ref{fig3}, we observe a qualitatively different picture. Cascades branching is visibly reduced as compared to low-energy events, in that we do not find well-separated sub-cascades. Some cascade branching is seen in the first 0.5 MeV cascade shown in Fig. \ref{fig3}a only and, importantly, during an intermediate stage of cascade development only. On the other hand, the final cascade morphology is described by a rather continuous distribution of the damage across about 1000-1300 \AA\ where no distinctly separated sub-cascades can be identified. Common to all collision cascades we have simulated, this picture is particularly visible for defect atoms in the final state of the cascades shown in \mbox{Figs. \ref{fig2}--\ref{fig3}}.

Qualitatively, reduced cascade branching and the emergence of a more continuous damage distribution can be understood as follows. For a scattered atom to move far enough from its initial position and form a spatially separated sub-cascade (i.e., branch out) requires a large value of energy transferred to it by the incident atom. In the absence of inelastic losses, the transferred energy, $T$, is $T=\frac{1}{2}T_m(1-\cos(\phi))$, where $T_m$ is the maximal transferred energy and $\phi$ is the scattering angle \cite{smith}. For large energy of the incident atom, $E$, $\phi$ decreases as $\phi\propto\frac{1}{E}$ \cite{smith}. We therefore find that for large $E$ and small $\phi$, $T$ decreases as $T\propto\phi^2\propto\frac{1}{E^2}$. Large $E$ and, consequently, small $T$, results in scattered atoms forming the damaged region close to the trajectory of the initial knock-on atom and, therefore, promote a continuous structure of the resulting damage. This is consistent with our current findings. 

We note that as the incident atom slows down, $T$ increases, leading to sub-cascade branching at the end of the atom trajectory. However, larger $E$ results in the increase of the relative fraction of continuous damage over the fraction of branched cascades.

Our finding is important in the context of the long-term evolution of radiation damage. Indeed, a recent kinetic Monte Carlo study \cite{bjork} has shown that very large defect clusters can have a major effect on the long-term damage development.

The discussion has so far concentrated on the large-scale cascade structure and morphology. We now briefly discuss defect structures at the local level. The size and structure of the defect clusters created by the cascades were analyzed and the results are summarized in \mbox{Table \ref{tab1}}. The simulations confirm that the normalized fraction of Frenkel pairs (FPs) of 0.3--0.4 is roughly constant for cascades over \mbox{0.1 MeV} \cite{malerba2, st-rev}. The number of surviving FPs follows the trend from lower cascade energies (Fig. 4). The fraction of surviving interstitials grouped into clusters was found to be 0.58(3) and 0.52(3) for cascade energies of 0.2 and 0.5 MeV, respectively. This is consistent with the results for 50-100 keV cascades \cite{st-rev, bjork} hinting at a possibility that the clustering fraction may reach a maximum at \mbox{$\sim $ 100 keV}. A similar trend was observed for vacancy clustering fractions of 0.33(3) and 0.35(1), respectively.

\begin{table*}[tb]
\begin{ruledtabular}
\setlength{\tabcolsep}{3pt}
\begin{tabular}{p{1.2cm} l p{2.0cm} p{1.8cm} p{1.8cm} p{1.8cm} p{1.8cm} p{1.5cm} p{1.8cm}}
    Cascade energy & $N_{\rm F}$ & NRT fraction of defects & Number of isolated vacancies & Number of split SIAs & Number of vacancy clusters & Number of SIA clusters & Largest vacancy cluster & Largest SIA~cluster \\
  \hline
  0.2 MeV            & 900 (200)  & 0.44 (0.11)  & 70 (5)      & 65 (4)      & 17 (1)      & 46 (7)  & 54 & 89 \\
  0.5 MeV            & 1450 (220) & 0.29 (0.04)  & 150 (14)    & 170 (15)    & 36 (7)      & 84 (13) & 47 & 36 \\
\end{tabular}
\end{ruledtabular}
\caption{The number of Frenkel pairs (FP), ($N_{\rm F}$), and defect statistics for 0.2 MeV and 0.5 MeV cascade simulations in $\alpha$-iron. The value in brackets shows standard error in the mean over 4 constituent runs for each simulation. Largest clusters are determined by net defect count. NRT fraction is the normalized number of FP \cite{nrt}.
}
\label{tab1}
\end{table*}

\begin{figure}[ht]
\begin{center}
\includegraphics[width=\columnwidth]{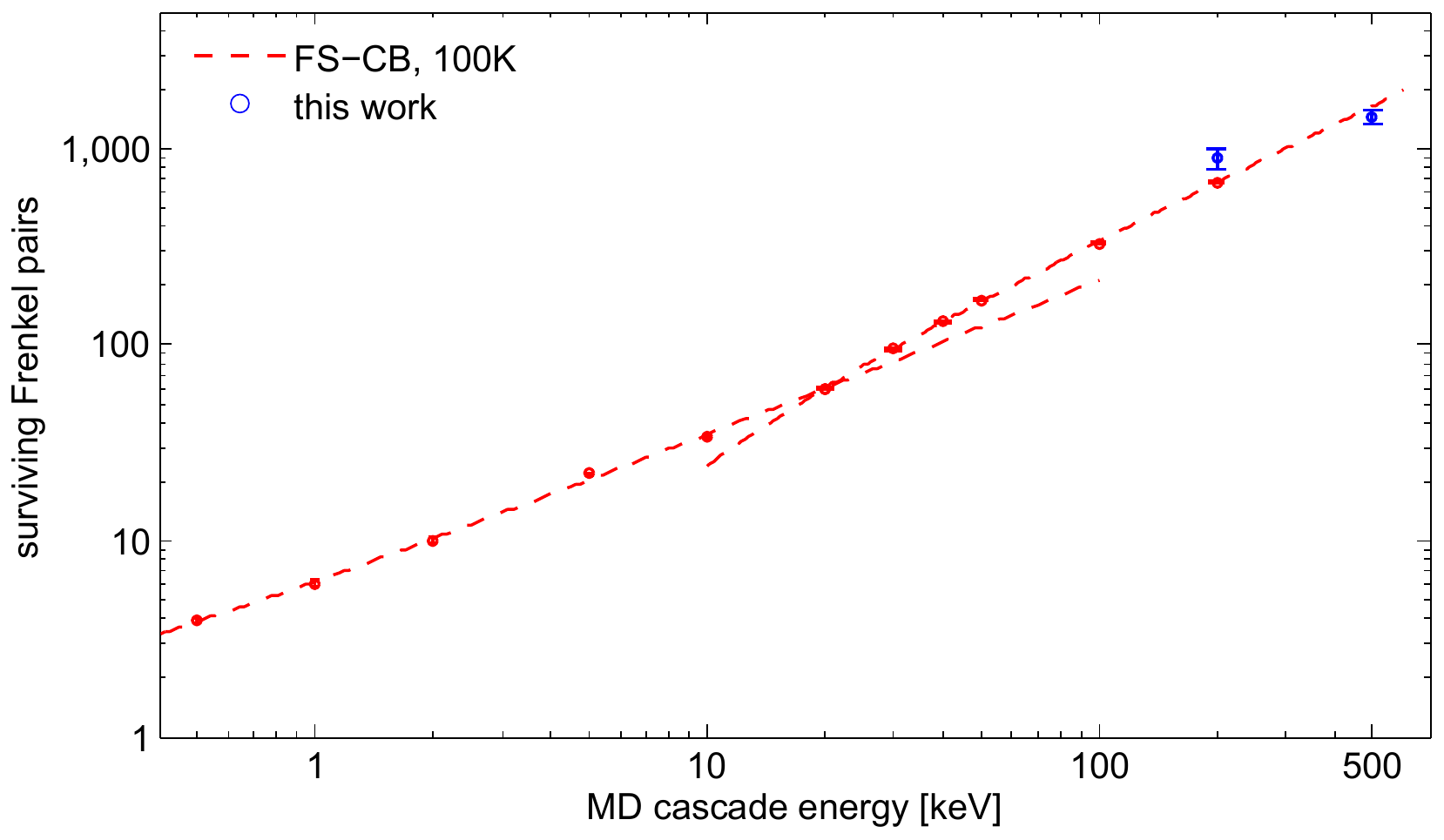}
\end{center}
\caption{
Dependence of the number of surviving FPs on the cascade energy. A comparison between Finnis-Sinclair potential runs started at 100 K (red line \cite{st-rev}) with the current work (blue). The fits correspond to a single-cascade and sub-cascades (branching) regimes (details in \cite{st-rev}).
}
\label{fig4}
\end{figure}

\begin{figure}[th]
\begin{center}
\includegraphics[width=\columnwidth]{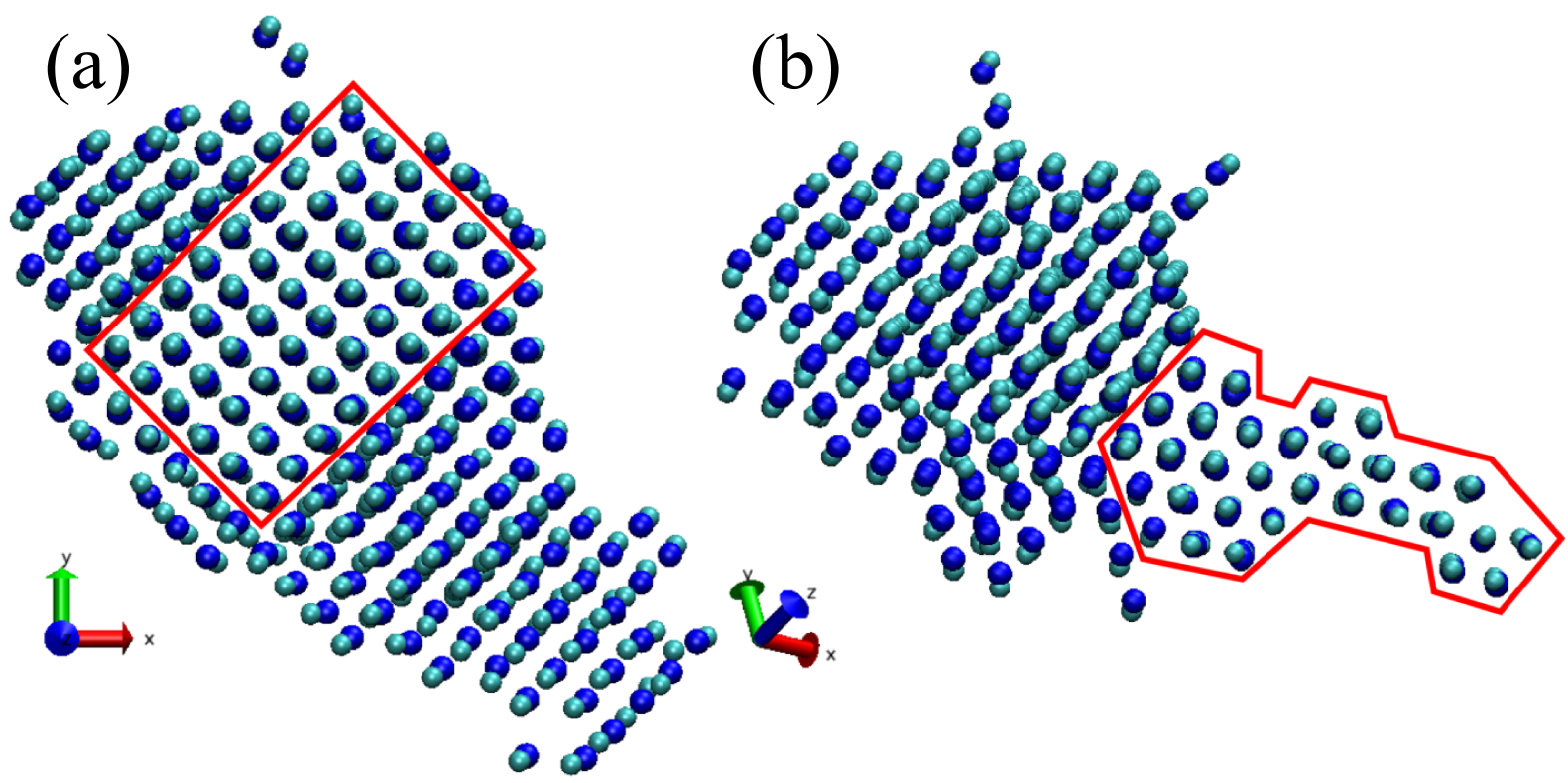}
\end{center}
\caption{
The largest cluster consisting of 89 intersitials. It is mainly composed of a set of $<$100$>$ crowdions (selected region in (a)) and a fraction of normally glissile $<$111$>$ crowdions (region highlighted in (b)). Such cluster morphology blocks the motion of crowdions and results in an overall sessile cluster; similar effect of immobilization of a cluster by another defect was observed in \cite{gao1}. Interstitials (vacancies) are shown in silver (blue). We used VMD package for visualization of local defect structures \cite{vmd}.
}
\label{f10}
\end{figure}

The majority of interstitial clusters were found to be glissile $<$111$>$ crowdion clusters. The largest defect structure was a composite 89-interstitial cluster, formed from a merger of a set of $<$111$>$ and $<$100$>$ crowdions \mbox{(Fig. \ref{f10})}. Owing to its complex morphology, it will be immobile. This interstitial cluster is quite large, yet consistent with the data reported for lower energies \cite{soneda,bjork}.

\begin{figure}[th]
\begin{center}
\includegraphics[width=\columnwidth]{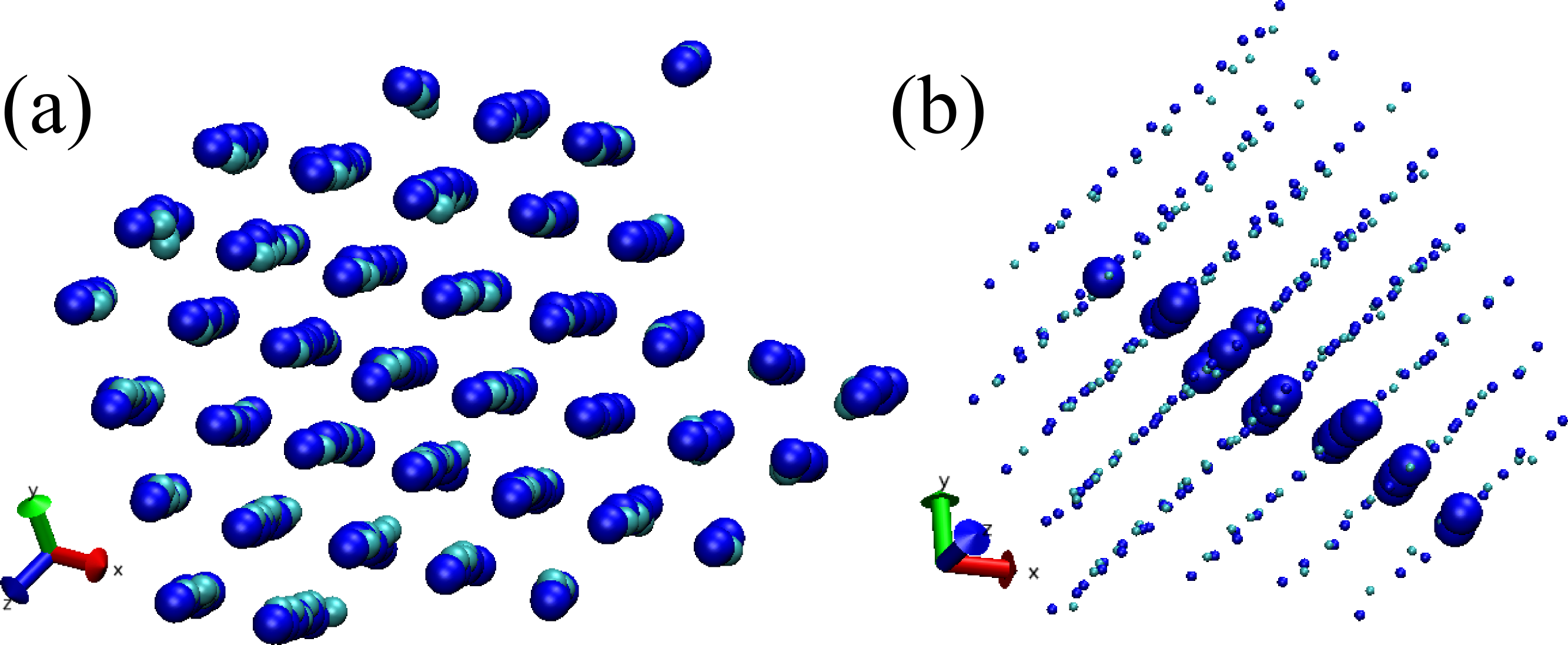}
\end{center}
\caption{
(a) A (111) projection of a 39-vacancy cluster; (b) a (001) projection of this cluster. The large spheres show the central vacancy for selected constituent `vacancy crowdions', thus emphasizing the dislocation nature of the cluster.
}
\label{f11}
\end{figure}

\begin{figure}[th]
\begin{center}
\includegraphics[width=\columnwidth]{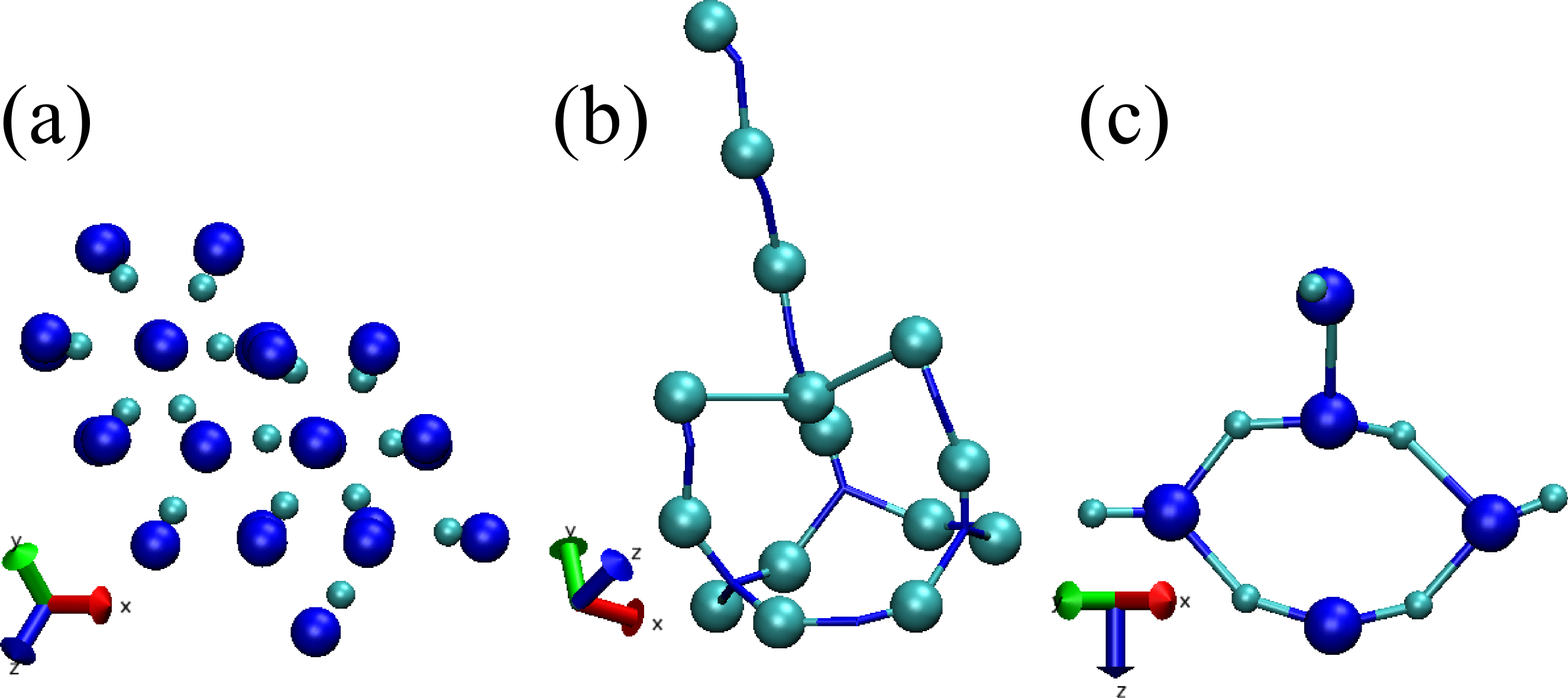}
\end{center}
\caption{(a) The (111) projection of a small 9-vacancy cluster; (b) a C15 phase tetra-interstitial with a $<$111$>$ crowdion attached. The vacancies are omitted from the figure for clarity; (c) a hexagonal \mbox{di-interstitial} with a split interstitial attached.
}
\label{f12}
\end{figure}

Large vacancy clusters were also observed with the 54-vacancy cluster being the largest one. Several large vacancy clusters formed $<$100$>$ and $<$111$>$ dislocation loop-like configurations [Fig. \ref{f11}(a)]. A cross section of an exemplar vacancy cluster is shown [Fig. \ref{f11}(b)], to emphasize its dislocation-like nature. The smaller vacancy clusters revealed a rich variety of structures, such as hexagonal vacancy clusters with interstitial rings surrounding a central vacancy [Fig. \ref{f12}(a)].

We also observed a wide range of sessile interstitial clusters. Some of these could be clearly identified as being related to the C15 meta-stable phase discussed in \cite{Marinica}, and many were joined to crowdions or crowdion clusters [Fig. \ref{f12}(b)]. Smaller ring-like structures were also observed [Fig. \ref{f12}(c)] in which 6 atoms shared 4 neighbouring lattice sites.

\section{Conclusion}

In summary, we have found novel structural features of radiation damage in iron on both large and local scale which we will need to be included in physical models aimed at understanding and predicting the effects of radiation damage on the mechanical, thermal and transport properties of structural materials. The reported damage structures such as the increased continuous morphology of high-energy collision cascades will form a starting point for long-timescale models in order to understand and predict the effects of radiation damage. Large defect structures reported here, including novel vacancy and interstitial clusters, will be important for understanding of the interaction of these clusters with transmutation gases and nucleation of helium bubbles.

\begin{acknowledgements}
Via our membership of the UK's HPC Materials Chemistry Consortium, which is funded by EPSRC (EP/F067496), this work made use of the facilities of HECToR, the UK's national high-performance computing service, which is provided by UoE HPCx Ltd at the University of Edinburgh, Cray Inc and NAG Ltd, and funded by the Office of Science and Technology through EPSRC's High End Computing Programme.
\end{acknowledgements}


\begin{thebibliography}{99}

\bibitem{dud1} D. J. Ward and S. L. Dudarev 2008 {\it Materials Today} {\bf 11} 46 

\bibitem{stoneham} A. M. Stoneham, J. R. Matthews and I. J. Ford 2004 {\it J. Phys.: Condens. Matter} {\bf 16} S2597

\bibitem{dud2} S. L. Dudarev et al 2009 {\it J. Nuclear Mater.} {\bf 1} 386

\bibitem{weber} W. J. Weber et al 1998 {\it J. Mater. Res.} {\bf 13} 1434

\bibitem{right} A. M. Stoneham and J. H. Harding 2003 {\it Nature Materials} {\bf 2} 77

\bibitem{synroc2} A. E. Ringwood et al 1981 {\it Nuclear and Chemical Waste Management} {\bf 2} 287 

\bibitem{gei1} T. Geisler, K. Trachenko, S. Rios, M. T. Dove, and E. K. H. Salje 2003 {\it J. Phys.: Condens. Matter} {\bf 15} L597 

\bibitem{Ave98} R. S. Averback and T. Diaz de la Rubia 1998 {\it Solid State Physics} (ed. H. Erhenfest and F. Spaepen, Academic Press, New York) {\bf 51} 281 

\bibitem{ishino} S. Ishino, P. Schiller and A. F. Rowcliffe 1989 {\it Journal of Fusion Energy} {\bf 8} 147

\bibitem{k6} K. Nordlund 2002 {\it Nuclear Instr. Meth. Phys. Res. B} {\bf 188} 41

\bibitem{k1} A. Souidi et al 2006 {\it J. Nucl. Mater.} {\bf 89} 355 

\bibitem{k2} A. F. Calder, D. J. Bacon, A. V. Barashev and Y. N. Osetsky 2008 {\it J. Nucl. Mater.} {\bf 382} 91 

\bibitem{k3} K. Morishita and T. Diaz de la Rubia 1999 {\it J. Nucl. Mater.} {\bf 271} 35 

\bibitem{k4} D. J. Bacon et al. 2003 {\it J. Nucl. Mater.} {\bf 152} 323 

\bibitem{stoller2} R. E. Stoller and L. R. Greenwood 1999 {\it Journal of Nuclear Materials} {\bf 271--272} 57 

\bibitem{st-rev} R. E. Stoller 2012 {\it Comprehensive Nuclear Materials} {\bf 1} 293 (Ed. R. J. M. Konings, Elsevier, Amsterdam, 2012).

\bibitem{dl1} I. T. Todorov, B. Smith, K. Trachenko and M. T. Dove 2005 {\it Capability Computing} {\bf 6} 12 

\bibitem{dl2} I. T. Todorov, B. Smith, M. T. Dove and K. Trachenko 2006 {\it J. Mater. Chem.} {\bf 16} 1911 

\bibitem{hector} http://www.hector.ac.uk

\bibitem{race} C. P. Race et al 2010 {\it Rep. Prog. Phys.} {\bf 73} 116501

\bibitem{duffy1} A. M. Rutherford and D. M. Duffy 2007 {\it J. Phys.: Condens. Mat.} {\bf 19} 496201 

\bibitem{caro} A. Caro, M. Victoria 1989 {\it Phys. Rev. A} {\bf 40} 2287

\bibitem{duffy2} D. M. Duffy, A. M. Rutherford 2009 {\it J. Nucl. Mater.} {\bf 386–-388} 19

\bibitem{mendel} M. I. Mendelev et al 2003 {\it Phil. Mag.} {\bf 83} 3977

\bibitem{malerba} L. Malerba et al 2010 {\it J. Nucl. Mat.} {\bf 406} 19 

\bibitem{zbl} J. F. Ziegler, J. P. Biersack, and U. Littmark 1985 {\it The Stopping and Range of Ions in Matter} (Pergamon, New York)

\bibitem{atomeye} J. Li 2003 {\it Modell. Simul. Mater. Sci. Eng.} {\bf 11} 173

\bibitem{spike} T. Diaz de la Rubia, R. S. Averback, R. Benedeck and W. E. King 1987 {\it Phys. Rev. Lett.} {\bf 59} 1930 

\bibitem{Sam07d} J. Samela and K. Nordlund 2008 {\it Phys. Rev. Lett.} {\bf 101} 027601 

\bibitem{prb} K. Trachenko, M. T. Dove, and E. Artacho, I. T. Todorov and W. Smith 2006 {\it Phys. Rev. B} {\bf 73} 174207

\bibitem{prb1} K. Trachenko, M. Pruneda, E. Artacho, and M. T. Dove 2005 {\it Phys. Rev. B} {\bf 71} 184104 

\bibitem{Nor98} K. Nordlund, L. Wei, Y. Zhong and R. S. Averback 1998 {\it Phys. Rev. B} {\bf 57} 13965

\bibitem{Hei94} H. L. Heinisch, B. N. Singh and T. Diaz de la Rubia 1994 {\it J. Nucl. Mat.} {\bf 212--215} 127 

\bibitem{smith} R. Smith 1997 {\it Atomic and ion collisions in solids and at surfaces} (Cambridge University Press)

\bibitem{soneda} N. Soneda, S. Ishino and T. Diaz de la Rubia 2001 {\it Phil. Mag. Lett.} {\bf 81} 649  

\bibitem{bjork} C. Bj\"{o}rkas K. Nordlund and M. J. Caturla 2012 {\it Phys. Rev. B} {\bf 85} 024105

\bibitem{nrt} M. J. Norgett, M.T. Robinson, I.M. Torrens 1975 {\it Nucl. Eng. Des.} {\bf 33} 50 

\bibitem{malerba2} L. Malerba 2006 {\it J. Nucl. Mat.} {\bf 351} 28 

\bibitem{gao1} F. Gao, D. J. Bacon, Y. N. Osetsky, P. E. J. Flewitt, and T. A. Lewis 2000 {\it J. Nucl. Mater.} {\bf 213-–220} 276

\bibitem{vmd} W. Humphrey, A. Dalke and K. Schulten 1996 {\it J. Molecular Graphics} {\bf 14} 33 

\bibitem{Marinica} M.-C. Marinica, F. Willaime, and J.-P. Crocombette 2012 {\it Phys. Rev. Lett.} {\bf 2} 108

\bibitem{ttm} S. I. Anisimov, B. L. Kapeliovich, and T. L. Perel'man 1974 {\it Zh. Eksp. Teor. Fiz.} {\bf 66} 776 [1974 {\it Sov. Phys. JETP} {\bf 39} 375].

\bibitem{lindhard} J. Lindhard and M. Scharff 1961 {\it Physical Review} {\bf 124} 1, 128–130

\bibitem{coh2} E. Zhurkin and A. Kolesnikov 2003 {\it Nucl. Instrum. Methods Phys. Res. Sect. B} {\bf 202} 269--277

\bibitem{coh1} M. Jakas and D. Harrison 1985 {\it Phys. Rev. B} {\bf 32} 2752--2760

\bibitem{coh3} K. Nordlund, L. Zhong, and R. Averback. 1998 {\it Phys. Rev. B} {\bf 57} 965--968

\bibitem{coh4} J. le Page, D. R. Mason, C. P. Race, and W. M. C. Foulkes 2009 {\it New Journal of Physics} {\bf 11} 1 013004

\bibitem{coh5} C. Bj\"{o}rkas, K. Nordlund 2009 {\it Nucl. Instrum. Methods Phys. Res. Sect. B} 267 {\bf 10} 1830--1836

\end{thebibliography}
\end{document}